\documentclass[twocolumn,showpacs,amssymb,aps,prl,floatfix]{revtex4}

\usepackage{latexsym} 	
\usepackage{amsmath}  	
\usepackage{amsbsy}
\usepackage{epsfig}

\newcommand{\om}{\omega}

\newcommand{\id}{1\!\!1}

\newcommand{\bra}{\langle}
\newcommand{\ket}{\rangle}

\newcommand{\half}{\frac{1}{2}}

\newcommand{\vecx}{{\mathbf x}}

\newcommand{\vecnul}{{\mathbf 0}}

\newcommand{\be}{\begin{equation}}
\newcommand{\ee}{\end{equation}}
\newcommand{\bea}{\begin{eqnarray}}
\newcommand{\eea}{\end{eqnarray}}
\newcommand{\bean}{\begin{eqnarray*}}
\newcommand{\eean}{\end{eqnarray*}}

\begin{document}

\title{Spectral functions at small energies and the electrical 
conductivity\\ in hot, quenched lattice QCD
}

\author{Gert Aarts}
\author{Chris Allton}
\author{Justin Foley}
\author{Simon Hands}

\affiliation{Department of Physics, Swansea University,
	Swansea SA2 8PP, United Kingdom}

\author{Seyong Kim}
\affiliation{Department of Physics, Sejong University,
        Seoul 143-747, Korea}

\date{March 12, 2007}

\begin{abstract}

In lattice QCD, the Maximum Entropy Method can be used to reconstruct 
spectral functions from euclidean correlators obtained in 
numerical simulations.  We show that at finite temperature the most 
commonly used algorithm, employing Bryan's method, is inherently unstable 
at small energies and give a modification that avoids this. We demonstrate 
this approach using the vector current-current correlator obtained
in quenched QCD at finite temperature. Our first results indicate a small 
electrical conductivity above the deconfinement transition.

\end{abstract}


\maketitle

%%%%%%%%%%%%%%%%%%%%%%%%%%%%%%%%%%%%%%%%%%%%%%%%%%%%%%%%%%%%%%%%%%%%%

% SECTION 

In the deconfined, high-temperature phase of Quantum Chromodynamics, the 
behaviour of spectral functions of conserved currents at small energies is 
of intrinsic interest due to its relation with transport properties of the 
quark-gluon plasma (QGP). According to the Kubo formulas \cite{Kadanoff}, 
transport coefficients, such as the shear and bulk viscosities and the 
electrical conductivity, are proportional to the slope of appropriate 
spectral functions at vanishing energy. The success of e.g.\ ideal 
hydrodynamics in heavy ion phenomenology, assuming vanishing viscosities 
and requiring early thermalization \cite{Heinz:2001xi}, has lead to the 
notion
 %\cite{Gyulassy:2004vg,Gyulassy:2004zy,Shuryak:2004cy}
 \cite{Gyulassy:2004vg}
 that the QGP created in relativistic heavy 
ion collisions at RHIC is strongly coupled and that the ratio of shear 
viscosity to entropy density in this sQGP may be close to the conjectured 
lower bound \cite{Kovtun:2004de} reached in thermal field theories that 
admit a gravity dual 
 %\cite{Policastro:2001yc,Buchel:2003tz}.
 \cite{Policastro:2001yc}.

 In order to put these ideas on firm footing, it is important to have a 
first-principle calculation of transport coefficients in the strongly 
coupled regime of hot QCD. As is well known 
\cite{Karsch:1986cq,Aarts:2002cc}, a nonperturbative calculation using 
lattice QCD is difficult due to the necessity to perform an analytic 
continuation from imaginary to real time. The most common approach used to 
obtain spectral functions from euclidean correlators is the Maximum 
Entropy Method \cite{Asakawa:2000tr}, employing Bryan's algorithm 
\cite{Bryan}. Our first aim in this Letter is to discuss this method in 
some detail and point out a source of numerical instabilities present in 
most finite-temperature studies available to date. We show how the 
algorithm can be modified to avoid this problem. Our second aim is to 
apply the new method to the vector current-current correlator, obtained in 
quenched lattice QCD simulations at finite temperature, using staggered 
quarks. We study the behaviour at small energies and argue that it allows 
us to extract a value for electrical conductivity in the strongly coupled 
regime above the deconfinement transition.

{\em Maximum Entropy Method} --
 The relation between the euclidean correlator $G(\tau) = \int d^3x \, 
\bra J(\tau,\vecx) J^\dagger(0,\vecnul) \ket$ at zero momentum and the 
corresponding spectral function $\rho(\om)$ reads
 \be
 \label{eqG}
 G(\tau) = \int_0^\infty \frac{d\omega}{2\pi} \, K(\om,\tau) \rho(\om),
\ee
 where the kernel is given by
\be
 K(\om,\tau) = \frac{\cosh[\om(\tau-1/2T)]}{\sinh(\om/2T)}.
\ee
  We consider (local) meson operators of the form $J(\tau,\vecx) = \bar 
q(\tau,\vecx) \Gamma q(\tau,\vecx)$, where $\Gamma$ depends on the channel 
under consideration.
 The temperature $T$ is related to the euclidean temporal extent $N_\tau$ 
by $1/T=a N_\tau$, where $a$ is the (temporal) lattice spacing.
 The difficulty in inverting relation (\ref{eqG}) is due to the fact that 
$G(\tau)$ is obtained numerically at a discrete set of points $\tau_i = 
\tau_{\rm min} + (i-1)a$ ($i=1,\ldots,N$), where the number of data points 
$N$ is typically ${\cal O}(10)$, whereas $\rho(\omega)$ is in principle a 
continuous function of $\om$. Simple properties of the kernel and spectral 
functions allow us to cutoff the $\omega$ integral at $\om_{\rm max}$. The 
resulting finite interval is discretized as $\om_n=n\Delta\omega$ 
($n=1,\ldots,N_\om$), where $N_\om$ is typically ${\cal O}(10^3)$, making 
a simple inversion ill-defined (below we suppress the index $n$ in $\om_n$ 
where possible).

Using the ideas of Bayesian probability theory, one may construct the most 
probable spectral function by maximizing the conditional probability 
$P[\rho|DH]$, where $D$ indicates the data and $H$ some additional prior 
knowledge. In the Maximum Entropy Method (MEM), the prior knowledge is 
encoded in an entropy term,
 \be
 S = \int_0^\infty \frac{d\om}{2\pi} \left[ \rho(\om) - m(\om) - 
\rho(\om)\ln\frac{\rho(\om)}{m(\om)} \right],
\ee
 where $m(\om)$ is the so-called default model, containing the additional 
information. An often used default model is $m(\om) = m_0\om^2$, with 
$m_0$ a (channel-dependent) constant. This choice is motivated by the 
large-$\om$ behaviour of meson spectral functions in the continuum theory, 
which is accessible in perturbation theory.
 The conditional probability to be extremized reads
 \be
 P[\rho|DH] = \exp \left(-\half \chi^2 +\alpha S\right),
\ee 
 where $\chi^2$ is the standard likelihood function and $\alpha$ is a 
parameter balancing the relative importance of the data and the prior 
knowledge. Since $\rho(\om)$ is nonnegative for positive $\om$, it is 
written as
 \be
\rho(\om) = m(\om) \exp f(\om),
\ee
 where $f(\om)$ is to be determined.

%%%%%%%%%%%%%%%%%%%%%%%%%%%%%%%%%%%%%%%%%%%%%%%%%%%%%%%%%

% SECTION 
                                                                     
{\em Bryan's approach} -- To make the problem well-defined, the arbitrary 
function $f(\om)$ has to be reduced to one containing at most $N$ 
parameters, i.e.\ it has to be restricted to an $N$ dimensional subspace.
 In Bryan's algorithm the subspace is determined using a singular value 
decomposition (SVD). After discretizing the $\omega$ integral, the 
kernel $K(\om_n,\tau_i)$ is viewed as a $N_\om\times N$ matrix. The SVD 
theorem states that it can be written as $K = U W V^T$, with $U$ an 
$N_\om\times N$ matrix satisfying $U^TU=\id_{N\times N}$, $W = 
\mbox{diag}(w_1,\dots, w_N)$ with $w_1\geq \ldots \geq w_N\geq 
0$, and $V$ an $N\times N$ matrix satisfying 
$VV^T=V^TV=\id_{N\times N}$. The dimension $N_s$ of the subspace is 
determined by the singular values $w_i$. Due to the reflection symmetry 
$K(\om,1/T-\tau) = K(\om,\tau)$, one finds that $N_s\leq N \leq 
N_\tau/2+1$. Discarding the contact term at $\tau=0$, we use below the 
maximal range $N_s=N=N_\tau/2$, such that all data points at $1\leq 
\tau_i/a\leq N_\tau/2$ are included.
 In Bryan's algorithm the subspace is defined as the space spanned by the 
column vectors of $U$, i.e.\ the $N$ vectors $u_i$ ($i=1,\ldots,N$) with 
elements $u_i(\om_n) = U_{ni}$. Since $U$ is orthogonal, these vectors are 
linearly independent and normalized, with the inner product $\bra 
u_i|u_j\ket \equiv \sum_{n=1}^{N_\om} u_i(\om_n)  u_j(\om_n) = 
\delta_{ij}$. It follows from the extremum condition 
that $f(\om)$ can be parametrized as 
 \be
 f(\om) = \sum_{i=1}^N c_i u_i(\om).
\ee
This reduces the problem to the determination of $N$ coefficients $c_i$, 
as desired.

\begin{figure}[t]
 \centerline{\epsfig{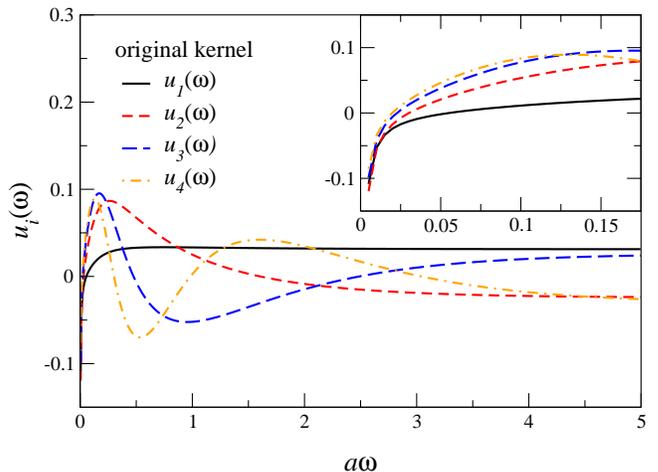}}
 \caption{First four basis functions $u_i(\om)$ as a 
function of $a\omega$ for $a\om_{\rm max}=5$, $N_\om=1000$, $N_\tau=24$, 
using the standard kernel. 
The inset shows a blow-up of the small energy region. 
}
 \label{fig:svd}
\end{figure}

The $u_i$ functions should therefore be regarded as basis functions and it 
is interesting to study how they behave. In Fig.\ \ref{fig:svd} we show 
the first four basis functions determined using the SVD of 
$K(\om_n,\tau_i)$ for the case that $N_\om=1000$ and $N=N_\tau/2=12$. The 
cutoff is $a\om_{\rm max}=5$. Details of the basis functions depend on the 
choice of $N_\om$ and $N$, but we find that the $i$'th function crosses 
zero $i$ times. The inset shows a blowup of the small energy behaviour. It 
can be seen that all basis functions seem to diverge in the small energy 
limit (although they are still normalizable). This is because the kernel 
itself diverges for small $\om$, since
 \be 
 \label{eqGsmall}
\lim_{\om\to 0} K(\om,\tau) = \frac{2T}{\om} + 
\frac{\om}{T}\left[\frac{1}{6}-\tau T(1-\tau T)\right] +{\cal 
O}\left(\frac{\om^3}{T^3}\right). 
 \ee
 Due to this divergence, it is not possible to include the point at 
$\om=0$ (note that the limits $T\to 0$ and $\om\to 0$ do not commute). 

In the MEM analysis of lattice correlators at finite temperature, it is 
often found (see e.g.\ Ref.\ \cite{Aarts:2006cq} and references therein) 
that the small energy behaviour is unstable: the value of the spectral 
function at the smallest nonzero $\om$ value is inconsistent with values 
at larger energies and the $\om\to 0$ behaviour depends on $\Delta\om$, 
indicating that it is an artefact of the method. Moreover MEM does not 
always converge. Unstable features in the algorithm at small energies 
prevent of course insight in transport properties of the QGP.

%%%%%%%%%%%%%%%%%%%%%%%%%%%%%%%%%%%%%%%%%%%%%%%%%%%%%%%%%%%%%

\begin{figure}[t]
 \centerline{\epsfig{figure=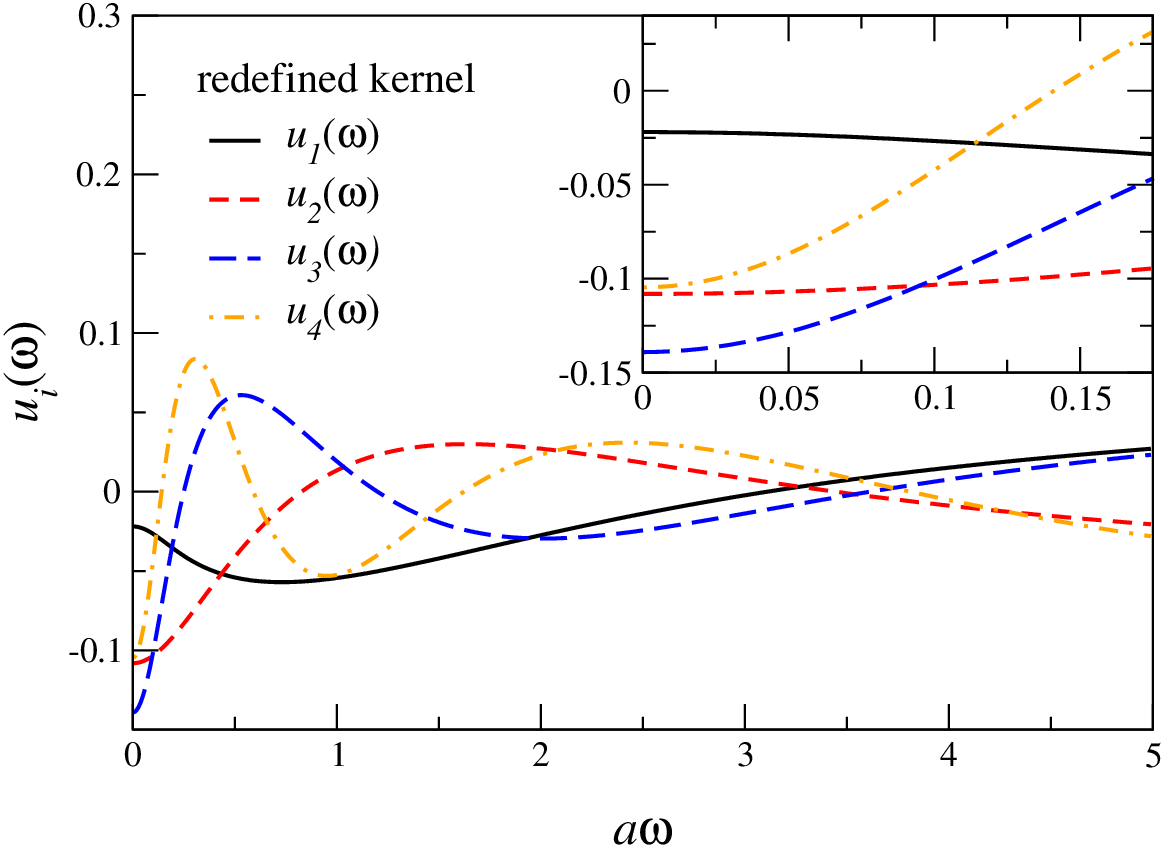,height=6.2cm}}
 \caption{As in Fig.\ \ref{fig:svd}, using the redefined kernel $\overline 
K(\om,\tau)$. 
}
 \label{fig:svd_mod}
\end{figure}

{\em Modification of Bryan's algorithm} --
 As indicated above, the divergence of the kernel at small energies can 
lead to a numerically unstable algorithm. Fortunately, this can easily be 
avoided by writing
 \be 
  \overline K(\om,\tau) = \frac{\om}{2T} K(\om,\tau),
\;\;\;\;\;\;\;\;
  \overline\rho(\om) = \frac{2T}{\om} \rho(\om),
\ee
 such that $ K(\om,\tau)\rho(\om) = \overline K(\om,\tau)\overline 
\rho(\om)$. We may now repeat the SVD for $\overline K(\om_n,\tau_i)$. We 
find that both kernels give the same identification of the dimension of 
the subspace. The first four new basis functions $\overline u_i$ are shown 
in Fig.\ \ref{fig:svd_mod}. The inset shows again a blow-up. Since 
$\overline K(0,\tau)=1$, the basis functions take a finite value at small 
$\om$ and $\overline u_i(0)$ is well-defined for all $i$.
 From now on we include the point at $\om=0$ in the analysis. 
 The redefined spectral function is parametrized as 
 \footnote{In Ref.\ 
\cite{Jakovac:2006sf} it is proposed to use $u_i(\om)=K(\om,\tau_i)$. 
Due to Eq.\ (\ref{eqGsmall}), this results in problematic small energy 
behaviour. One way to avoid this is by using $\overline{K}(\om,\tau_i)$. 
}
  \be
 \label{eqrhobar}
  \overline\rho(\om) = \overline m(\om) \exp \sum_{i=1}^N \overline c_i 
\overline u_i(\om),
 \ee
 with the default model $\overline m(\om) \sim m(\om)/\om$. In order to 
make contact with previous results, $\overline m(\om)\sim \om$ at large 
$\om$. For small $\om$, we note that we reconstruct 
$\overline\rho\sim\rho/\om$, such that the intercept at $\om=0$ is 
proportional to the appropriate transport coefficient. Specifically, for 
the electrical conductivity we find that
 \be
 \frac{\sigma}{T} = \lim_{\om\to 0}\, \frac{\rho(\om)}{6\om T} = 
 \frac{\overline\rho(0)}{12T^2},
 \ee
 where in this case $\rho$ is in the vector channel ($\Gamma=\gamma_k$, 
summed over $k=1,2,3$). To allow for a nonzero intercept, we find from 
Eq.\ (\ref{eqrhobar}) that $\overline m(0)$ should be finite 
and nonzero. We use therefore the following default model
 \be
 \label{eqdefault}
 a^2\overline m(\om) = \overline{m}_0 (b+a\om),
 \ee
 where $b\lesssim 1$ is a parameter that can be varied to assess the 
default-model dependence of spectral functions in the low-energy regime.

%%%%%%%%%%%%%%%%%%%%%%%%%%%%%%%%%%%%%%%%%%%%%%%%%%%%%%%%%

% SECTION 

\begin{table}[b]
\begin{center}
\begin{tabular}{|l|c|c|c|c|c|}
\hline
     & 	$\beta$ & $a^{-1}$ (GeV) & $N_\sigma^3\times N_\tau$ & $T/T_c$ & \# conf \\  
\hline
cold &  6.5	& 4.04	& $48^3\times 24$	& 0.62	 & 100 \\
hot  &  7.192   & 9.72 	& $64^3\times 24$      	& 1.5    & 100 \\
very hot & 7.192& 9.72	& $64^3\times 16$      	& 2.25   & 50 \\
\hline
\end{tabular}
 \caption{
 Lattice parameters. Estimates for the lattice spacing and temperature are 
taken from Ref.\ \cite{Datta:2003ww}.
 }
\label{table1}
\end{center}
\end{table}

{\em Lattice QCD} -- We now apply the modified algorithm to the euclidean 
correlator in the vector channel, obtained in quenched lattice QCD 
simulations at finite temperature. Lattice details are given in Table 
\ref{table1}. The two lattices above $T_c$ differ only in temperature, 
while the lattice below $T_c$ is coarser but has $N_\tau=24$, in common 
with the hot one. We use light staggered quarks with a bare mass of 
$am=0.01$.
 As is well known, the pure gauge action has a global $Z_{3}$ symmetry, 
which is broken by the presence of fermions in full QCD. To incorporate 
this, we multiply the link variables by an element of $Z_{3}$ in the 
calculation of the quark propagators so that the phase of the Polyakov 
loop is approximately real. Chiral symmetry restoration at $T>T_c$ is then 
clearly visible: the pseudoscalar and scalar correlators coincide 
\cite{Aarts:2006cq}.

For staggered fermions the MEM analysis is complicated due to the mixing 
of two signals in the correlator. The equivalent of relation (\ref{eqG}) 
reads
 \be
 \label{eqGstag}
 G^{\rm stag}(\tau) = \int_0^\infty \frac{d\om}{2\pi}\, K(\tau,\om)
\left[
\rho(\om) - (-1)^{\tau/a}\tilde \rho(\om) \right],
\ee
 where $\tilde \rho$ is related to $\rho$ by changing $\Gamma$ to
$\tilde\Gamma = \gamma_4\gamma_5\Gamma$. In practice we perform an
independent MEM analysis on the even and odd time slices, obtaining
$\rho_{\rm even} = \rho-\tilde \rho$ and $\rho_{\rm odd} =
\rho+\tilde \rho$ and combine these to get $\rho$.
 Let us remark here that while the original formulation, using 
$K(\om,\tau)$, failed to converge in some cases, we found that the new 
method worked successfully (we have studied the vector and pseudoscalar 
channels \cite{inprep}). In cases where both methods worked we found 
comparable results for spectral functions in most of the energy range but 
deviations in the low-energy region, as expected.

\begin{figure}[t]
 \centerline{\epsfig{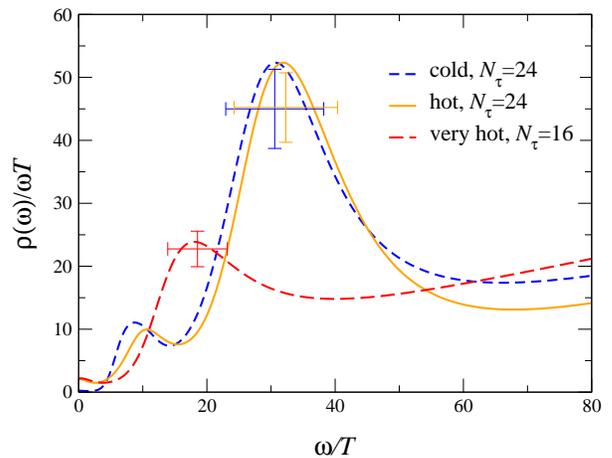}}
 \caption{ 
 Vector spectral functions $\rho(\om)/\om T$ as a function of $\om/T$. We 
used $N_\om=1000$, $a\om_{\rm max}=5$, $b=1$.
 }
 \label{fig:rho}
\end{figure}

In Fig.\ \ref{fig:rho} we show the vector spectral functions 
$\rho(\om)$ normalized by $\om T$ as a function of $\om/T$ for the 
three temperatures. 
 Our treatment of $\alpha$ follows Ref.\ \cite{Asakawa:2000tr}. 
 The large $\om$ behaviour at $a\om\gtrsim 3$ is determined by the 
continuum default model, $\rho(\om)/\om \sim m_0\om$ (recall that 
$\om/T=N_\tau a\om$).
 In most MEM studies we carried out we find ``bumps'' at $1\lesssim a\om 
\lesssim 2$: within uncertainties these do not depend on the channel under 
consideration, quark mass or external momentum \cite{Aarts:2006cq,inprep}. 
We interpret them as lattice artefacts present due to the difference 
between continuum and lattice fermion dispersion relation (see Ref.\ 
\cite{Aarts:2005hg} for an analysis of free staggered quarks). The peak 
structure at $\om/T\sim 9$ on the cold lattice is the rho-meson, which is 
more clearly visible in plots of $\rho(\om)/\om^2$ vs.\ $\om/T$ 
\cite{Aarts:2006cq}. The rho-peak is less pronounced 
on the hot lattice and seems to have vanished on the very hot lattice. 
Error bars are explained in Ref.~\cite{Asakawa:2000tr}.

\begin{figure}[t]
 \centerline{\epsfig{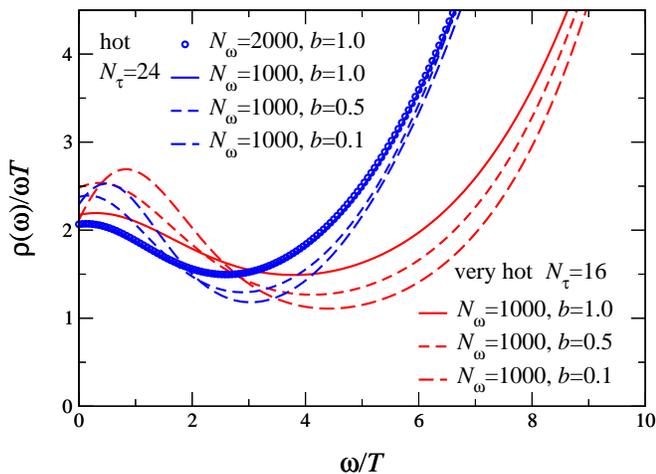}}
 \caption{
 Default model dependence of $\rho(\om)/\om T$ for $N_\tau=24$ (hot)
and $16$ (very hot) in the low-energy region. We show 
results for $N_\om=1000, 2000$ and $b=1.0, 0.5, 0.1$ at fixed $a\om_{\rm 
max}=5$. 
 }
 \label{fig:rho_zoom}
\end{figure}

Above $T_c$ a nonzero intercept at $\om=0$ can be seen. In the cold phase 
the intercept is an order of magnitude smaller. A blow-up of 
$\rho(\om)/\om T$ is shown in Fig.\ \ref{fig:rho_zoom}.
 In the high-temperature weakly-coupled theory, the transport contribution 
is located at $\om/T\sim g^4(T) \ll 1$, where the spectral function 
behaves as $\rho(\om)/\om T \sim 1/g^4(T) \gg 1$. As a result the 
euclidean correlator is insensitive to the details of the spectral 
function \cite{Aarts:2002cc}.
 Here, by contrast, we find a clear signal for nonzero spectral weight 
varying smoothly in the range $0\lesssim \om/T \lesssim 4$.
 To assess this, we have varied $N_\om$, $a\om_{\rm max}$ and the 
parameter $b$ in Eq.\ (\ref{eqdefault}). In the case that we take $b=0$ 
(therefore disallowing a nonzero intercept), we find that the MEM routine 
does not converge. Results with $N_\om=2000$ cannot be distinguished from 
those obtained with $N_\om=1000$. These findings suggest that nonzero 
spectral weight and a finite intercept are robust outcomes.
 Dividing the intercept at $\om=0$ by 6 yields the electrical 
conductivity. We find $\sigma/T = 0.4 \pm 0.1$, where the error is an 
indication of the uncertainty in the MEM reconstruction 
\cite{Gupta:2003zh} \footnote{This result is normalized to a single 
flavour and should be multiplied with the sum of the electric charge 
squared for light flavours.}. Statistical uncertainty and the effect of 
operator normalization \cite{Daniel:1987aa} are expected to be 
smaller.

%%%%%%%%%%%%%%%%%%%%%%%%%%%%%%%%%%%%%%%%%%%%%%%%%%%%%%%%%

% SECTION 

{\em Outlook} -- We identified a source for numerical instabilities 
in the standard formulation of MEM with Bryan's approach at finite 
temperature and showed how this can be avoided. 
 We then applied this new method to the vector current-current correlator 
obtained in quenched lattice QCD at finite temperature. We found a robust 
signal for a small but nonzero value of the electrical conductivity above 
the deconfinement transition, in line with the notion of the sQGP. 
 Directions for the future are many. It is important to repeat the 
analysis using Wilson fermions and the conserved point split current, and 
to include the shear and bulk viscosity as well. The effect of dynamical 
quarks can be investigated in two-flavour QCD on highly anisotropic 
lattices \cite{Aarts:2007pk}.
 We emphasize once more that the new formulation is essential for 
analyzing the low-energy region.

%***************************************************************

\begin{acknowledgments}

 We thank Steve Sharpe, Ulrich Heinz, Antal Jakovac and Peter Petreczky
 for useful email correspondence. 
 G.A.\ is supported by a PPARC Advanced Fellowship. 
 S.K.\ is supported partly by a PPARC Visiting Fellowship and by the Korea 
Research Foundation Grant funded by the Korean Government (MOEHRD, Basic 
Research Promotion Fund) (KRF-2006-C00020).

\end{acknowledgments}

%***************************************************************

\end{document}